\documentclass[a4paper,11pt]{article}
\pdfoutput=1 

\usepackage[T1]{fontenc} 
\usepackage{booktabs}
\usepackage{amsfonts}
\usepackage{tikz}
\usetikzlibrary{shapes,arrows,chains,decorations.pathreplacing}

\usepackage{subfig}

\usepackage{alphalph}

\usepackage{hyperref}
\hypersetup{colorlinks=true,urlcolor=blue,citecolor=blue}
\usepackage{xcolor}
\usepackage{amsmath}
\definecolor{blu}{rgb}{0.,0.,1.}
\definecolor{red}{rgb}{1.,0.,0.}
\definecolor{burgundy}{rgb}{0.5, 0.0, 0.13}
\definecolor{crimsonred}{rgb}{0.6, 0.0, 0.0}
\definecolor{persianblue}{rgb}{0.11, 0.22, 0.73}
\definecolor{forestgreen}{rgb}{0.13,0.35,0.13}

\usepackage[backend=biber,sorting=none,giveninits=true,isbn=false,bibstyle=na62eb]{biblatex}
\title{\boldmath }

\newcommand{\RNum}[1]{\expandafter{\romannumeral #1\relax}}

\begin{document} 
\pagenumbering{arabic}
\centerline{\LARGE EUROPEAN ORGANIZATION FOR NUCLEAR RESEARCH}
\begin{flushright}
CERN-EP-2023-066\\
14 April 2023
\end{flushright}
\vspace{15mm}

\begin{center}
\Large{\bf Improved calorimetric particle identification in NA62 using machine learning techniques\\
\vspace{5mm}
}

\end{center}

\begin{center}
{\Large The NA62 Collaboration}\\
\end{center}
\vspace{10mm}

\begin{abstract}
Measurement of the ultra-rare \mbox{$K^+\to\pi^+\nu\bar\nu$} decay at the NA62 experiment at CERN requires high-performance particle identification to distinguish muons from pions. Calorimetric identification currently in use, based on a boosted decision tree algorithm, achieves a muon misidentification probability of $1.2\times 10^{-5}$ for a pion identification efficiency of 75\% in the momentum range of 15--40~GeV/$c$. In this work, calorimetric identification performance is improved by developing an algorithm based on a convolutional neural network classifier augmented by a filter. Muon misidentification probability is reduced by a factor of six with respect to the current value for a fixed pion-identification efficiency of 75\%. Alternatively, pion identification efficiency is improved from 72\% to 91\% for a fixed muon misidentification probability of $10^{-5}$.
\end{abstract}

\newcommand{\orcimg}{\raisebox{-0.3\height}{\includegraphics[height=\fontcharht\font`A]{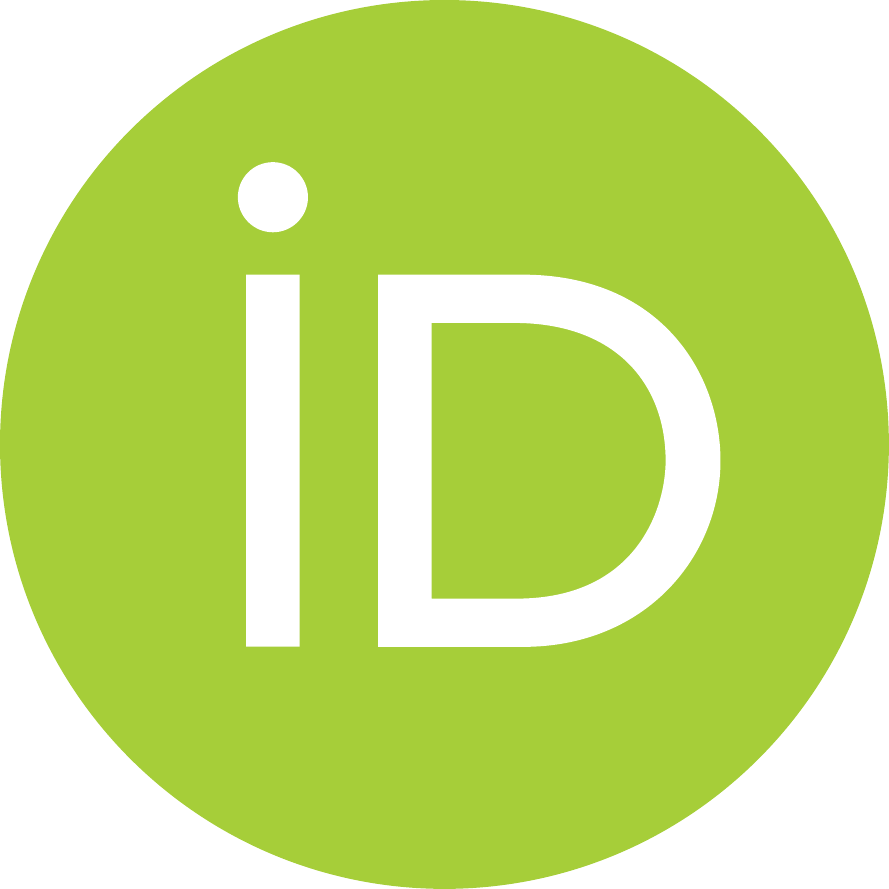}}}
\newcommand{\orcid}[1]{\href{https://orcid.org/#1}{\orcimg}}

\centerline{\bf The NA62 Collaboration} 
\vspace{0.5cm}
%
%

\begin{raggedright}
\noindent
{\bf Universit\'e Catholique de Louvain, Louvain-La-Neuve, Belgium}\\
 E.~Cortina Gil\orcid{0000-0001-9627-699X},
 A.~Kleimenova$\,${\footnotemark[1]}\orcid{0000-0002-9129-4985},
 E.~Minucci$\,${\footnotemark[2]}\orcid{0000-0002-3972-6824},
 S.~Padolski$\,${\footnotemark[3]}\orcid{0000-0002-6795-7670},
 P.~Petrov, 
 A.~Shaikhiev$\,${\footnotemark[4]}\orcid{0000-0003-2921-8743},
 R.~Volpe$\,${\footnotemark[5]}\orcid{0000-0003-1782-2978}
\vspace{0.5cm}

{\bf TRIUMF, Vancouver, British Columbia, Canada}\\
 W.~Fedorko\orcid{0000-0002-5138-3473},
 T.~Numao\orcid{0000-0001-5232-6190},
 Y.~Petrov\orcid{0000-0003-2643-8740},
 B.~Velghe$\,$\renewcommand{\thefootnote}{\fnsymbol{footnote}}\footnotemark[1]\renewcommand{\thefootnote}{\arabic{footnote}}\orcid{0000-0002-0797-8381},
 V. W. S.~Wong\orcid{0000-0001-5975-8164},
 M.~Yu\orcid{0000-0002-7387-9250}
\vspace{0.5cm}

{\bf University of British Columbia, Vancouver, British Columbia, Canada}\\
 D.~Bryman$\,${\footnotemark[6]}\orcid{0000-0002-9691-0775},
 J.~Fu
\vspace{0.5cm}

{\bf Charles University, Prague, Czech Republic}\\
 Z.~Hives\orcid{0000-0002-5025-993X},
 T.~Husek$\,${\footnotemark[7]}\orcid{0000-0002-7208-9150},
 J.~Jerhot$\,${\footnotemark[8]}\orcid{0000-0002-3236-1471},
 K.~Kampf\orcid{0000-0003-1096-667X},
 M.~Zamkovsky$\,${\footnotemark[9]}\orcid{0000-0002-5067-4789}
\vspace{0.5cm}

{\bf Institut f\"ur Physik and PRISMA Cluster of Excellence, Universit\"at Mainz, Mainz, Germany}\\
  A.T.~Akmete\orcid{0000-0002-5580-5477},
 R.~Aliberti$\,${\footnotemark[10]}\orcid{0000-0003-3500-4012},
 G.~Khoriauli$\,${\footnotemark[11]}\orcid{0000-0002-6353-8452},
 J.~Kunze,
 D.~Lomidze$\,${\footnotemark[12]}\orcid{0000-0003-3936-6942},
 L.~Peruzzo\orcid{0000-0002-4752-6160}, 
 M.~Vormstein,
 R.~Wanke\orcid{0000-0002-3636-360X}
\vspace{0.5cm}

{\bf Dipartimento di Fisica e Scienze della Terra dell'Universit\`a e INFN, Sezione di Ferrara, Ferrara, Italy}\\
 P.~Dalpiaz,
 M.~Fiorini\orcid{0000-0001-6559-2084},
 I.~Neri\orcid{0000-0002-9047-9822},
 A.~Norton$\,${\footnotemark[13]}\orcid{0000-0001-5959-5879},
 F.~Petrucci\orcid{0000-0002-7220-6919},
 M.~Soldani\orcid{0000-0003-4902-943X},
 H.~Wahl$\,${\footnotemark[14]}\orcid{0000-0003-0354-2465}
\vspace{0.5cm}

{\bf INFN, Sezione di Ferrara, Ferrara, Italy}\\
 L.~Bandiera\orcid{0000-0002-5537-9674},
 A.~Cotta Ramusino\orcid{0000-0003-1727-2478},
 A.~Gianoli\orcid{0000-0002-2456-8667}
 A.~Mazzolari\orcid{0000-0003-0804-6778},
 M.~Romagnoni\orcid{0000-0002-2775-6903},
 A.~Sytov\orcid{0000-0001-8789-2440}
\vspace{0.5cm}

{\bf Dipartimento di Fisica e Astronomia dell'Universit\`a e INFN, Sezione di Firenze, Sesto Fiorentino, Italy}\\
 E.~Iacopini\orcid{0000-0002-5605-2497},
 G.~Latino\orcid{0000-0002-4098-3502},
 M.~Lenti\orcid{0000-0002-2765-3955},
 P.~Lo Chiatto\orcid{0000-0002-4177-557X},
 I.~Panichi\orcid{0000-0001-7749-7914},
 A.~Parenti\orcid{0000-0002-6132-5680}
\vspace{0.5cm}

{\bf INFN, Sezione di Firenze, Sesto Fiorentino, Italy}\\
 A.~Bizzeti$\,${\footnotemark[15]}\orcid{0000-0001-5729-5530},
 F.~Bucci\orcid{0000-0003-1726-3838}
\vspace{0.5cm}

{\bf Laboratori Nazionali di Frascati, Frascati, Italy}\\
 A.~Antonelli\orcid{0000-0001-7671-7890},
 G.~Georgiev$\,${\footnotemark[16]}\orcid{0000-0001-6884-3942},
 V.~Kozhuharov$\,${\footnotemark[16]}\orcid{0000-0002-0669-7799},
 G.~Lanfranchi\orcid{0000-0002-9467-8001},
 S.~Martellotti\orcid{0000-0002-4363-7816}, 
 M.~Moulson\orcid{0000-0002-3951-4389},
 T.~Spadaro\orcid{0000-0002-7101-2389},
 G.~Tinti\orcid{0000-0003-1364-844X}
\vspace{0.5cm}

{\bf Dipartimento di Fisica ``Ettore Pancini'' e INFN, Sezione di Napoli, Napoli, Italy}\\
 F.~Ambrosino\orcid{0000-0001-5577-1820},
 T.~Capussela,
 M.~Corvino\orcid{0000-0002-2401-412X},
 M.~D'Errico\orcid{0000-0001-5326-1106},
 D.~Di Filippo\orcid{0000-0003-1567-6786},
 R.~Fiorenza$\,${\footnotemark[17]}\orcid{0000-0003-4965-7073}, 
 R.~Giordano\orcid{0000-0002-5496-7247},
 P.~Massarotti\orcid{0000-0002-9335-9690},
 M.~Mirra\orcid{0000-0002-1190-2961},
 M.~Napolitano\orcid{0000-0003-1074-9552},
 G.~Saracino\orcid{0000-0002-0714-5777}
\vspace{0.5cm}

{\bf Dipartimento di Fisica e Geologia dell'Universit\`a e INFN, Sezione di Perugia, Perugia, Italy}\\
 G.~Anzivino\orcid{0000-0002-5967-0952},
 F.~Brizioli$\,${\footnotemark[9]}\orcid{0000-0002-2047-441X},
 E.~Imbergamo,
 R.~Lollini\orcid{0000-0003-3898-7464},
 R.~Piandani$\,${\footnotemark[18]}\orcid{0000-0003-2226-8924},
 C.~Santoni\orcid{0000-0001-7023-7116}
\vspace{0.5cm}

{\bf INFN, Sezione di Perugia, Perugia, Italy}\\
 M.~Barbanera\orcid{0000-0002-3616-3341},
 P.~Cenci\orcid{0000-0001-6149-2676},
 B.~Checcucci\orcid{0000-0002-6464-1099},
 P.~Lubrano\orcid{0000-0003-0221-4806},
 M.~Lupi$\,${\footnotemark[19]}\orcid{0000-0001-9770-6197}, 
 M.~Pepe\orcid{0000-0001-5624-4010},
 M.~Piccini\orcid{0000-0001-8659-4409}
\vspace{0.5cm}

{\bf Dipartimento di Fisica dell'Universit\`a e INFN, Sezione di Pisa, Pisa, Italy}\\
 F.~Costantini\orcid{0000-0002-2974-0067},
 L.~Di Lella$\,${\footnotemark[14]}\orcid{0000-0003-3697-1098},
 N.~Doble$\,${\footnotemark[14]}\orcid{0000-0002-0174-5608},
 M.~Giorgi\orcid{0000-0001-9571-6260},
 S.~Giudici\orcid{0000-0003-3423-7981}, 
 G.~Lamanna\orcid{0000-0001-7452-8498},
 E.~Lari\orcid{0000-0003-3303-0524},
 E.~Pedreschi\orcid{0000-0001-7631-3933},
 M.~Sozzi\orcid{0000-0002-2923-1465}
\vspace{0.5cm}

{\bf INFN, Sezione di Pisa, Pisa, Italy}\\
 C.~Cerri,
 R.~Fantechi\orcid{0000-0002-6243-5726},
 L.~Pontisso$\,${\footnotemark[20]}\orcid{0000-0001-7137-5254},
 F.~Spinella\orcid{0000-0002-9607-7920}
\vspace{0.5cm}

{\bf Scuola Normale Superiore e INFN, Sezione di Pisa, Pisa, Italy}\\
 I.~Mannelli\orcid{0000-0003-0445-7422}
\vspace{0.5cm}

{\bf Dipartimento di Fisica, Sapienza Universit\`a di Roma e INFN, Sezione di Roma I, Roma, Italy}\\
 G.~D'Agostini\orcid{0000-0002-6245-875X},
 M.~Raggi\orcid{0000-0002-7448-9481}
\vspace{0.5cm}

{\bf INFN, Sezione di Roma I, Roma, Italy}\\
 A.~Biagioni\orcid{0000-0001-5820-1209},
 P.~Cretaro\orcid{0000-0002-2229-149X},
 O.~Frezza\orcid{0000-0001-8277-1877},
 E.~Leonardi\orcid{0000-0001-8728-7582},
 A.~Lonardo\orcid{0000-0002-5909-6508}, 
 M.~Turisini\orcid{0000-0002-5422-1891},
 P.~Valente\orcid{0000-0002-5413-0068},
 P.~Vicini\orcid{0000-0002-4379-4563}
\vspace{0.5cm}

{\bf INFN, Sezione di Roma Tor Vergata, Roma, Italy}\\
 R.~Ammendola\orcid{0000-0003-4501-3289},
 V.~Bonaiuto$\,${\footnotemark[21]}\orcid{0000-0002-2328-4793},
 A.~Fucci,
 A.~Salamon\orcid{0000-0002-8438-8983},
 F.~Sargeni$\,${\footnotemark[22]}\orcid{0000-0002-0131-236X}
\vspace{0.5cm}

{\bf Dipartimento di Fisica dell'Universit\`a e INFN, Sezione di Torino, Torino, Italy}\\
 R.~Arcidiacono$\,${\footnotemark[23]}\orcid{0000-0001-5904-142X},
 B.~Bloch-Devaux\orcid{0000-0002-2463-1232},
 M.~Boretto$\,${\footnotemark[9]}\orcid{0000-0001-5012-4480},
 E.~Menichetti\orcid{0000-0001-7143-8200},
 E.~Migliore\orcid{0000-0002-2271-5192},
 D.~Soldi\orcid{0000-0001-9059-4831}
\vspace{0.5cm}

{\bf INFN, Sezione di Torino, Torino, Italy}\\
 C.~Biino\orcid{0000-0002-1397-7246},
 A.~Filippi\orcid{0000-0003-4715-8748},
 F.~Marchetto\orcid{0000-0002-5623-8494}
\vspace{0.5cm}

{\bf Instituto de F\'isica, Universidad Aut\'onoma de San Luis Potos\'i, San Luis Potos\'i, Mexico}\\
 A.~Briano Olvera\orcid{0000-0001-6121-3905},
 J.~Engelfried\orcid{0000-0001-5478-0602},
 N.~Estrada-Tristan$\,${\footnotemark[24]}\orcid{0000-0003-2977-9380}
 M. A.~Reyes Santos$\,${\footnotemark[24]}\orcid{0000-0003-1347-2579}
\vspace{0.5cm}

{\bf Horia Hulubei National Institute for R\&D in Physics and Nuclear Engineering, Bucharest-Magurele, Romania}\\
 P.~Boboc\orcid{0000-0001-5532-4887},
 A. M.~Bragadireanu,
 S. A.~Ghinescu\orcid{0000-0003-3716-9857},
 O. E.~Hutanu
\vspace{0.5cm}

{\bf Faculty of Mathematics, Physics and Informatics, Comenius University, Bratislava, Slovakia}\\
 L.~Bician$\,${\footnotemark[25]}\orcid{0000-0001-9318-0116},
 T.~Blazek\orcid{0000-0002-2645-0283},
 V.~Cerny\orcid{0000-0003-1998-3441},
 Z.~Kucerova$\,${\footnotemark[9]}\orcid{0000-0001-8906-3902}
\vspace{0.5cm}

{\bf CERN, European Organization for Nuclear Research, Geneva, Switzerland}\\
 J.~Bernhard\orcid{0000-0001-9256-971X},
 A.~Ceccucci\orcid{0000-0002-9506-866X},
 H.~Danielsson\orcid{0000-0002-1016-5576},
 N.~De Simone$\,${\footnotemark[26]},
 F.~Duval, 
 B.~D\"obrich$\,${\footnotemark[27]}\orcid{0000-0002-6008-8601},
 L.~Federici\orcid{0000-0002-3401-9522},
 E.~Gamberini\orcid{0000-0002-6040-4985},
 L.~Gatignon$\,${\footnotemark[28]}\orcid{0000-0001-6439-2945},
 R.~Guida, 
 F.~Hahn$\,$\renewcommand{\thefootnote}{\fnsymbol{footnote}}\footnotemark[2]\renewcommand{\thefootnote}{\arabic{footnote}},
 E. B.~Holzer\orcid{0000-0003-2622-6844},
 B.~Jenninger,
 M.~Koval$\,${\footnotemark[25]}\orcid{0000-0002-6027-317X},
 P.~Laycock$\,${\footnotemark[3]}\orcid{0000-0002-8572-5339}, 
 G.~Lehmann Miotto\orcid{0000-0001-9045-7853},
 P.~Lichard\orcid{0000-0003-2223-9373},
 A.~Mapelli\orcid{0000-0002-4128-1019},
 R.~Marchevski$\,${\footnotemark[1]}\orcid{0000-0003-3410-0918},
 K.~Massri$\,${\footnotemark[28]}\orcid{0000-0001-7533-6295}, 
 M.~Noy,
 V.~Palladino\orcid{0000-0002-9786-9620},
 M.~Perrin-Terrin$\,${\footnotemark[29]}$^,$$\,${\footnotemark[30]}\orcid{0000-0002-3568-1956},
 J.~Pinzino$\,${\footnotemark[31]}\orcid{0000-0002-7418-0636},
 V.~Ryjov, 
 S.~Schuchmann\orcid{0000-0002-8088-4226},
 S.~Venditti
\vspace{0.5cm}
\newpage
{\bf School of Physics and Astronomy, University of Birmingham, Birmingham, United Kingdom}\\
 T.~Bache\orcid{0000-0003-4520-830X},
 M. B.~Brunetti$\,${\footnotemark[32]}\orcid{0000-0003-1639-3577},
 V.~Duk$\,${\footnotemark[5]}\orcid{0000-0001-6440-0087},
 V.~Fascianelli$\,${\footnotemark[33]},
 J. R.~Fry\orcid{0000-0002-3680-361X}, 
 F.~Gonnella\orcid{0000-0003-0885-1654},
 E.~Goudzovski\orcid{0000-0001-9398-4237},
 J.~Henshaw\orcid{0000-0001-7059-421X},
 L.~Iacobuzio,
 C.~Lazzeroni\orcid{0000-0003-4074-4787}, 
 N.~Lurkin$\,${\footnotemark[8]}\orcid{0000-0002-9440-5927},
 F.~Newson,
 C.~Parkinson\orcid{0000-0003-0344-7361},
 A.~Romano\orcid{0000-0003-1779-9122},
 J.~Sanders\orcid{0000-0003-1014-094X}, 
 A.~Sergi$\,${\footnotemark[34]}\orcid{0000-0001-9495-6115}, 
 A.~Sturgess\orcid{0000-0002-8104-5571},
 J.~Swallow$\,${\footnotemark[9]}\orcid{0000-0002-1521-0911},
 A.~Tomczak\orcid{0000-0001-5635-3567}
\vspace{0.5cm}

{\bf School of Physics, University of Bristol, Bristol, United Kingdom}\\
 H.~Heath\orcid{0000-0001-6576-9740},
 R.~Page,
 S.~Trilov\orcid{0000-0003-0267-6402}
\vspace{0.5cm}

{\bf School of Physics and Astronomy, University of Glasgow, Glasgow, United Kingdom}\\
 B.~Angelucci,
 D.~Britton\orcid{0000-0001-9998-4342},
 C.~Graham\orcid{0000-0001-9121-460X},
 D.~Protopopescu\orcid{0000-0002-3964-3930}
\vspace{0.5cm}

{\bf Faculty of Science and Technology, University of Lancaster, Lancaster, United Kingdom}\\
 J.~Carmignani$\,${\footnotemark[35]}\orcid{0000-0002-1705-1061},
 J. B.~Dainton,
 R. W. L.~Jones\orcid{0000-0002-6427-3513},
 G.~Ruggiero$\,${\footnotemark[36]}\orcid{0000-0001-6605-4739}
\vspace{0.5cm}

{\bf School of Physical Sciences, University of Liverpool, Liverpool, United Kingdom}\\
 L.~Fulton,
 D.~Hutchcroft\orcid{0000-0002-4174-6509},
 E.~Maurice$\,${\footnotemark[37]}\orcid{0000-0002-7366-4364},
 B.~Wrona\orcid{0000-0002-1555-0262}
\vspace{0.5cm}

{\bf Physics and Astronomy Department, George Mason University, Fairfax, Virginia, USA}\\
 A.~Conovaloff,
 P.~Cooper,
 D.~Coward$\,${\footnotemark[38]}\orcid{0000-0001-7588-1779},
 P.~Rubin\orcid{0000-0001-6678-4985}
\vspace{0.5cm}

{\bf Authors affiliated with an Institute or an international laboratory covered by a cooperation agreement with CERN}\\
 A.~Baeva,
 D.~Baigarashev$\,${\footnotemark[39]}\orcid{0000-0001-6101-317X},
 D.~Emelyanov,
 T.~Enik\orcid{0000-0002-2761-9730},
 V.~Falaleev$\,${\footnotemark[5]}\orcid{0000-0003-3150-2196}, 
 S.~Fedotov,
 K.~Gorshanov\orcid{0000-0001-7912-5962},
 E.~Gushchin\orcid{0000-0001-8857-1665},
 V.~Kekelidze\orcid{0000-0001-8122-5065},
 D.~Kereibay, 
 S.~Kholodenko\orcid{0000-0002-0260-6570},
 A.~Khotyantsev, 
 A.~Korotkova,
 Y.~Kudenko\orcid{0000-0003-3204-9426},
 V.~Kurochka,
 V.~Kurshetsov\orcid{0000-0003-0174-7336},
 L.~Litov$\,${\footnotemark[16]}\orcid{0000-0002-8511-6883}, 
 D.~Madigozhin\orcid{0000-0001-8524-3455},
 M.~Medvedeva,
 A.~Mefodev,
 M.~Misheva$\,${\footnotemark[40]},
 N.~Molokanova, 
 S.~Movchan,
 V.~Obraztsov\orcid{0000-0002-0994-3641},
 A.~Okhotnikov\orcid{0000-0003-1404-3522}, 
 A.~Ostankov$\,$\renewcommand{\thefootnote}{\fnsymbol{footnote}}\footnotemark[2]\renewcommand{\thefootnote}{\arabic{footnote}},
 I.~Polenkevich,
 Yu.~Potrebenikov, 
 A.~Sadovskiy\orcid{0000-0002-4448-6845},
 V.~Semenov$\,$\renewcommand{\thefootnote}{\fnsymbol{footnote}}\footnotemark[2]\renewcommand{\thefootnote}{\arabic{footnote}},
 S.~Shkarovskiy,
 V.~Sugonyaev\orcid{0000-0003-4449-9993},
 O.~Yushchenko\orcid{0000-0003-4236-5115},
 A.~Zinchenko$\,$\renewcommand{\thefootnote}{\fnsymbol{footnote}}\footnotemark[2]\renewcommand{\thefootnote}{\arabic{footnote}}\orcid{0000-0002-1544-3395}
\vspace{0.5cm}

\end{raggedright}

%
%

\setcounter{footnote}{0}
\newlength{\basefootnotesep}
\setlength{\basefootnotesep}{\footnotesep}

\renewcommand{\thefootnote}{\fnsymbol{footnote}}
\noindent
$^{\footnotemark[1]}${Corresponding author: B.~Velghe, email: bvelghe@triumf.ca}\\
$^{\footnotemark[2]}${Deceased}\\
\renewcommand{\thefootnote}{\arabic{footnote}}
$^{1}${Present address: Ecole Polytechnique F\'ed\'erale Lausanne, CH-1015 Lausanne, Switzerland} \\
$^{2}${Present address: Syracuse University, Syracuse, NY 13244, USA} \\
$^{3}${Present address: Brookhaven National Laboratory, Upton, NY 11973, USA} \\
$^{4}${Present address: School of Physics and Astronomy, University of Birmingham, Birmingham, B15 2TT, UK} \\
$^{5}${Present address: INFN, Sezione di Perugia, I-06100 Perugia, Italy} \\
$^{6}${Also at TRIUMF, Vancouver, British Columbia, V6T 2A3, Canada} \\
$^{7}${Present address: Department of Astronomy and Theoretical Physics, Lund University, Lund, SE 223-62, Sweden} \\
$^{8}${Present address: Universit\'e Catholique de Louvain, B-1348 Louvain-La-Neuve, Belgium} \\
$^{9}${Present address: CERN, European Organization for Nuclear Research, CH-1211 Geneva 23, Switzerland} \\
$^{10}${Present address: Institut f\"ur Kernphysik and Helmholtz Institute Mainz, Universit\"at Mainz, Mainz, D-55099, Germany} \\
$^{11}${Present address: Universit\"at W\"urzburg, D-97070 W\"urzburg, Germany} \\
$^{12}${Present address: European XFEL GmbH, D-22761 Hamburg, Germany} \\
$^{13}${Present address: School of Physics and Astronomy, University of Glasgow, Glasgow, G12 8QQ, UK} \\
$^{14}${Present address: Institut f\"ur Physik and PRISMA Cluster of Excellence, Universit\"at Mainz, D-55099 Mainz, Germany} \\
$^{15}${Also at Dipartimento di Scienze Fisiche, Informatiche e Matematiche, Universit\`a di Modena e Reggio Emilia, I-41125 Modena, Italy} \\
$^{16}${Also at Faculty of Physics, University of Sofia, BG-1164 Sofia, Bulgaria} \\
$^{17}${Present address: Scuola Superiore Meridionale e INFN, Sezione di Napoli, I-80138 Napoli, Italy} \\
$^{18}${Present address: Instituto de F\'isica, Universidad Aut\'onoma de San Luis Potos\'i, 78240 San Luis Potos\'i, Mexico} \\
$^{19}${Present address: Institut am Fachbereich Informatik und Mathematik, Goethe Universit\"at, D-60323 Frankfurt am Main, Germany} \\
$^{20}${Present address: INFN, Sezione di Roma I, I-00185 Roma, Italy} \\
$^{21}${Also at Department of Industrial Engineering, University of Roma Tor Vergata, I-00173 Roma, Italy} \\
$^{22}${Also at Department of Electronic Engineering, University of Roma Tor Vergata, I-00173 Roma, Italy} \\
$^{23}${Also at Universit\`a degli Studi del Piemonte Orientale, I-13100 Vercelli, Italy} \\
$^{24}${Also at Universidad de Guanajuato, 36000 Guanajuato, Mexico} \\
$^{25}${Present address: Charles University, 116 36 Prague 1, Czech Republic} \\
$^{26}${Present address: DESY, D-15738 Zeuthen, Germany} \\
$^{27}${Present address: Max-Planck-Institut f\"ur Physik (Werner-Heisenberg-Institut), M\"unchen, D-80805, Germany} \\
$^{28}${Present address: Faculty of Science and Technology, University of Lancaster, Lancaster, LA1 4YW, UK} \\
$^{29}${Present address: Aix Marseille University, CNRS/IN2P3, CPPM, F-13288, Marseille, France} \\
$^{30}${Also at Universit\'e Catholique de Louvain, B-1348 Louvain-La-Neuve, Belgium} \\
$^{31}${Present address: INFN, Sezione di Pisa, I-56100 Pisa, Italy} \\
$^{32}${Present address: Department of Physics, University of Warwick, Coventry, CV4 7AL, UK} \\
$^{33}${Present address: Center for theoretical neuroscience, Columbia University, New York, NY 10027, USA} \\
$^{34}${Present address: Dipartimento di Fisica dell'Universit\`a e INFN, Sezione di Genova, I-16146 Genova, Italy} \\
$^{35}${Present address: School of Physical Sciences, University of Liverpool, Liverpool, L69 7ZE, UK} \\
$^{36}${Present address: Dipartimento di Fisica e Astronomia dell'Universit\`a e INFN, Sezione di Firenze, I-50019 Sesto Fiorentino, Italy} \\
$^{37}${Present address: Laboratoire Leprince Ringuet, F-91120 Palaiseau, France} \\
$^{38}${Also at SLAC National Accelerator Laboratory, Stanford University, Menlo Park, CA 94025, USA} \\
$^{39}${Also at L.N. Gumilyov Eurasian National University, 010000 Nur-Sultan, Kazakhstan} \\
$^{40}${Present address: Institute of Nuclear Research and Nuclear Energy of Bulgarian Academy of Science (INRNE-BAS), BG-1784 Sofia, Bulgaria} \\

\clearpage

\section{Introduction}
\label{sec:intro}
The NA62 experiment at CERN is dedicated to measurements of charged kaon decays, including the highly-suppressed ``golden mode'' \mbox{$K^+\to\pi^+\nu\bar\nu$}~\cite{CortinaGil2021} with a Standard Model branching ratio of $(8.4\pm 1.0)\times 10^{-11}$~\cite{Buras2015}. Background suppression requires a $\mu^+$ misidentification probability below $10^{-7}$, which is achieved by a combination of calorimetric and Cherenkov $\pi^+$ identification.

Machine learning (ML) methods developed primarily for image recognition can be applied to particle identification (PID) at NA62. Convolutional neural networks (CNN), a type of deep neural network, have been successfully used for classification tasks commonly encountered in particle physics~\cite{Baldi2014,deOliveira2016,Baldi2016}.
In this work, a calorimetric PID algorithm based on ML models is designed for the NA62 experiment. Two ML models are considered: LightGBM~\cite{ke2017lightgbm}, which is an implementation of a gradient boosting machine (GBM)~\cite{friedman2001greedy}, and a CNN-based model. Performance of the algorithm is compared to that of the boosted decision tree (BDT) implemented using TMVA GradientBoost~\cite{Hoecker2007}, currently used by NA62.

\section{NA62 beam and detector}
\label{sec:det}
\begin{figure}[t]
\includegraphics[width=\textwidth]{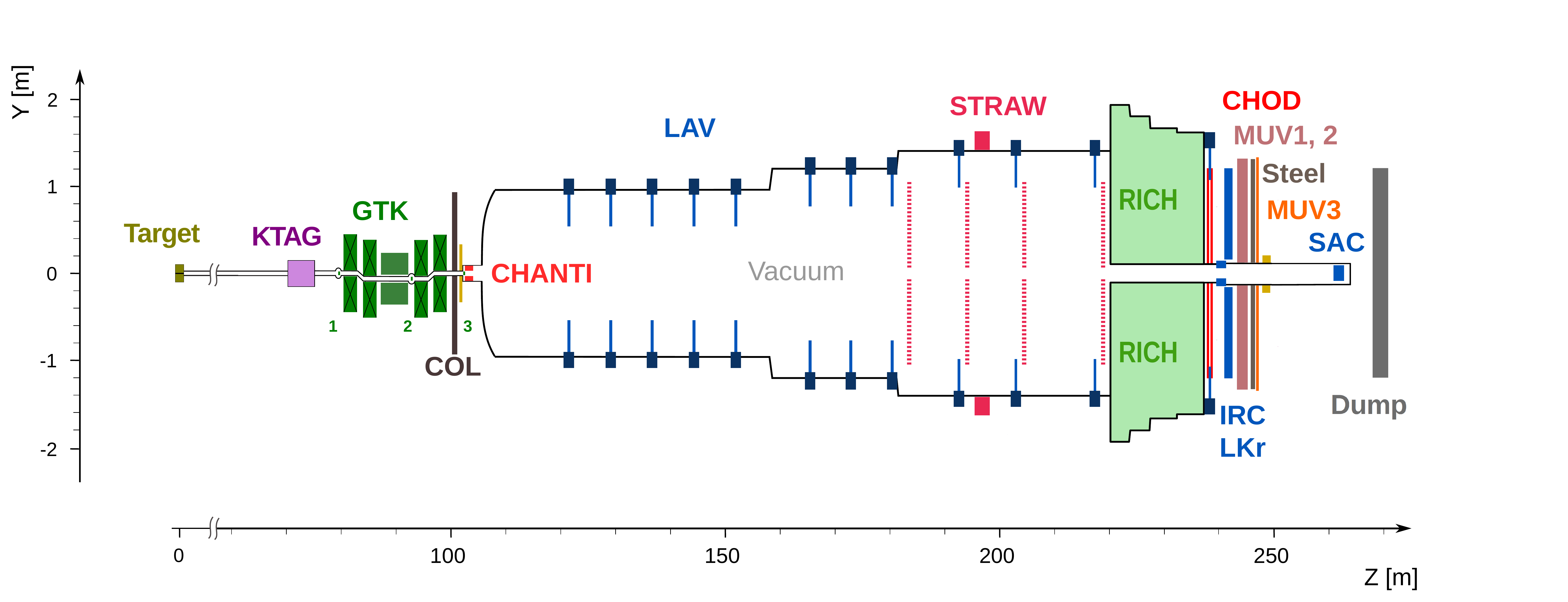}
\vspace{-8mm}
\caption{Schematic side view of the NA62 detector used in 2018.}
\label{fig:detector}
\end{figure}

The layout of the NA62 detector~\cite{Gil_2017} is shown schematically in Fig.~\ref{fig:detector}.
An unseparated secondary beam of $\pi^+$ (70\%), protons (23\%), and $K^+$ (6\%) is created by directing 400~GeV/$c$ protons extracted from the CERN SPS onto a beryllium target in spills of 3~s effective duration.
The central beam momentum is 75~GeV/$c$, and the momentum spread is 1\% (rms).
Beam kaons are tagged with 70~ps time resolution by a differential Cherenkov counter (KTAG) using a nitrogen gas radiator at 1.75~bar pressure contained in a 5~m long vessel. Beam particle positions, momenta and times are measured by a silicon pixel spectrometer consisting of three stations (GTK1, 2, 3) and four dipole magnets. A 1.2~m thick steel collimator (COL) is placed upstream of GTK3 to absorb hadrons produced in upstream $K^+$ decays. Inelastic interactions of beam particles in the GTK are detected by an array of scintillator hodoscopes (CHANTI). The beam is delivered into a vacuum tank evacuated to a pressure of $10^{-6}$~mbar, which contains a 75~m long fiducial decay volume (FV) starting 2.6~m downstream of GTK3.

Downstream of the FV, undecayed beam particles continue their path in vacuum. Momenta of charged particles produced in $K^+$ decays in the FV are measured by a magnetic spectrometer (STRAW) located in the vacuum tank downstream of the FV.
The spectrometer consists of four tracking chambers made of straw tubes, and a dipole magnet located between the the second and third chambers that provides a horizontal momentum kick of 270~MeV/$c$.
The spectrometer momentum resolution is $\sigma_p/p= (0.30 \oplus 0.005 \cdot p)\%$, where the momentum $p$ is expressed in~GeV/$c$.

A ring-imaging Cherenkov detector (RICH), consisting of a 17.5~m long vessel filled with neon at atmospheric pressure (with a Cherenkov threshold for pions of 12.5~GeV/$c$), is used for the identification of charged particles and for time measurement with 70~ps precision.
Two scintillator hodoscopes (CHOD) located downstream of the RICH provide trigger signals and time measurements with 200~ps precision.

A quasi-homogeneous liquid-krypton (LKr) electromagnetic calorimeter of thickness 27~radiation lengths is used for particle identification and photon detection.
The calorimeter has an active volume of 7~m$^3$, is segmented in the transverse plane into 13248 projective cells of $2\times 2$~cm$^2$, and 
provides an energy resolution $\sigma_E/E=(4.8/\sqrt{E}\oplus 11/E \oplus 0.9)\%$, where $E$ is expressed in GeV.
To achieve hermetic acceptance for photons emitted in the FV by $K^+$ decays at angles up to 50~mrad to the beam axis, the LKr calorimeter is supplemented by annular lead glass detectors (LAV) installed in 12~positions inside and downstream of the vacuum tank, and two lead/scintillator sampling calorimeters (IRC, SAC) located close to the beam axis.

A steel/scintillator hadronic sampling calorimeter is formed of two modules (MUV1, 2) with transverse dimensions of approximately $2.6\times 2.6$~m$^2$. The MUV1~(MUV2) module consists of 24~steel plates of 26.8~(25)~mm thickness, interleaved with 9~(4.5)~mm thick scintillator layers, resulting in a total thickness of 4.1~(3.7) interaction lengths. The scintillator layers in both modules are made of strips aligned alternately in the horizontal and vertical directions in consecutive planes. Each MUV1 scintillator layer consists of 48~strips, including 40~strips spanning the entire transverse length of the detector (read out at both ends), and 8~central strips of half-length (read out at the peripheral end), leading to $2\times 44$ channels per plane. Each MUV2 scintillator layer consists of $2\times 22$~strips, each spanning half the transverse length and read out at the peripheral end. In both modules, all strips located in the same transverse position in different layers are coupled to the same photomultiplier, thus providing a horizontal and a vertical view. A muon detector (MUV3) is located behind a 80~cm (4.8 interaction lengths) thick iron wall, has a transverse size of $2.64\times 2.64$~m$^2$, and is built from 50~mm thick scintillator tiles, including 140 regular tiles of $220 \times 220$~mm$^2$ transverse dimensions, and eight smaller central tiles adjacent to the beam pipe.

A two-level trigger system~\cite{AMMENDOLA20191,trig_na62_2022} is employed to reduce the event rate from 10~MHz to about 100~kHz. Multiple trigger lines with different downscaling factors are operated concurrently.

\section{GBM and CNN algorithms}
Supervised ML algorithms are used to construct a function (called {\it model} or {\it classifier}), $f(x_i;\theta)$, mapping a set of inputs $x_i$ (called {\it examples}) consisting of a number of attributes (called {\it features}) to a set of outputs $y_i$ (called {\it labels}). 
The set of model parameters, $\theta$, is chosen to minimise a {\it loss function} (also called {\it objective function}), $\mathcal{L} = \sum_i \ell\left(y_i,f(x_i;\theta)\right)$, where the function $\ell\left(y,f(x;\theta)\right)$ is selected depending on the problem to solve~\cite{Goodfellow2016,Murphy2022}. 

GBM is a generic family of algorithms used to build strongly predictive classifiers as linear combinations of {\it weak} classifiers (such as decision trees). The classifier, $f(x_i)$, is constructed iteratively, starting from an initial decision tree. 
At each step $m$, residuals are computed using the loss function obtained at the previous step as $-\partial \ell\left(y_i,f_{m-1}(x_i) \right) / \partial f_{m-1}$, and a regression tree fitting the residuals is constructed
and added to the linear combination.
This procedure is repeated until a stopping condition is met~\cite{friedman2001greedy}.

CNN algorithms have been developed to address computer vision tasks such as translation-invariant image classification, and can recognize complex geometric features at multiple scales. CNNs consist of consecutive \textit{layers}, the output of each layer being the input to the next one. Each layer is represented by a function $g(x)$, where $x$ is an input tensor.
Common layer types include convolutional, batch normalisation and pooling layers~\cite{Goodfellow2016,Murphy2022}. 
It has been observed~\cite{he2016deep} that increasing the number of layers leads to degradation of the training performance; to mitigate this issue, the ResNet architecture~\cite{he2016deep} introduces groups of consecutive layers called the \textit{residual blocks}. Instead of learning the parameters of the underlying function $g(x)$ directly, the block learns the parameters of the residual function $g_r(x)$, encoding the difference between the input and output of the layer. Many of these blocks are typically stacked.

\section{Data samples}
\label{sec:data}
Two independent datasets are employed: a {\it training and validation dataset} is used to train the models and adjust the hyperparameters (i.e. the parameters not derived via training), while an independent {\it test dataset} is used to characterise the performance of the optimal model with a frozen set of hyperparameters. The training and validation (test) dataset is based on 2.5\% (1.5\%) of the data sample collected by the NA62 experiment in 2016--2018. The training and validation dataset contains events collected with a non-downscaled $K^+\to \pi^+\nu\bar\nu$ trigger line~\cite{CortinaGil2021} based on RICH and CHOD signals in the absence of in-time MUV3 signals, and a control trigger line based on the CHOD signal downscaled by a factor of~400. The test dataset contains events collected with the control trigger line. This provides sufficient statistics for training, ensuring that the test dataset is free from trigger-induced bias.

High-purity pion ($\pi^+$), muon ($\mu^+$) and positron ($e^+$) track samples in the momentum range 15--50~GeV/$c$ are obtained from $K^+\to\pi^+ \pi^0$, $K^+\to\mu^+\nu$ and $K^+\to\pi^0 e^+\nu$ decay candidates selected without calorimetric particle identification, following the procedure used earlier for the existing NA62 BDT algorithm. Isolated STRAW tracks with associated in-time KTAG, CHOD, RICH and LKr signals and a single associated GTK track are considered. The decay vertex, reconstructed as the point of closest approach of the STRAW and GTK tracks, is required to be located in the FV and within the beam envelope.
For the $K^+\to\pi^+\pi^0$ and $K^+\to\pi^0 e^+\nu$ decays, a prompt $\pi^0\to\gamma\gamma$ decay is reconstructed by measuring the photons in the LKr calorimeter. Further selection criteria are based on photon veto conditions, particle identification in the RICH, and missing mass squared, $m_\mathrm{miss}^2 = (P_K - P_\mathrm{tr})^2$, where $P_K$ and $P_\mathrm{tr}$ are the $K^+$ and decay track four-momenta, respectively.  A label ($\pi^+$, $\mu^+$ or $e^+$) is assigned to each selected track.

The following features are encoded as matrices and vectors: track momentum and impact position in the LKr front plane; presence of a matching MUV3 signal; energy deposits in a matrix of $22 \times 22$ LKr calorimeter cells centered around the track impact position (these matrices are sparse due to the LKr Moli\`ere radius of 4.7~cm); total LKr energy deposit associated to the track; energy deposits in the horizontal and vertical views in all MUV1 channels ($2 \times 44$ channels in each view) and all MUV2 channels ($2\times 22$ channels in each view). Energy deposits within 10~ns of the track time are considered; the energy deposit values in every channel or cell are divided by the track momentum.

\begin{figure}[tb]
\begin{center}%
\includegraphics[width=0.92\textwidth]{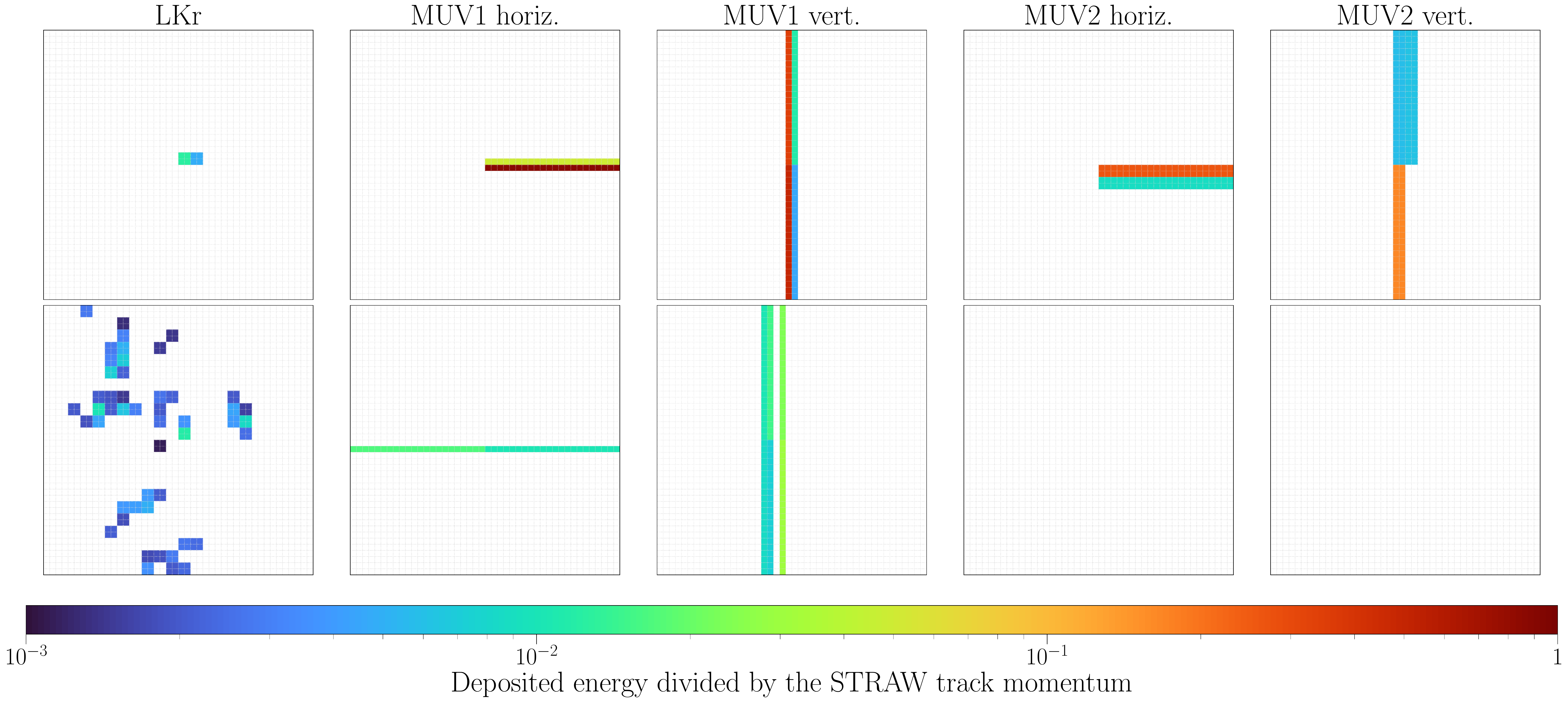}
\end{center}%
\vspace{-13mm}
\caption{CNN input matrices of $44\times 44$ size constructed for a $\mu^+$ track (top row) and a $\pi^+$ track (bottom row) from the training dataset. The LKr matrix is centred at the track impact point in the LKr front plane. In the MUV matrices related to the horizontal (vertical) view, the vertical (horizontal) coordinate is centered at the track impact point in the LKr front plane.}
\label{fig:event_display}
\end{figure}

The GBM algorithms require input information in the form of a single vector for each track. Therefore the LKr, MUV1 and MUV2 energy deposit matrices are transformed into vectors;
the LKr matrix is cropped to a size of $18 \times 18$ to exploit its sparsity. The results are concatenated into a 588-element feature vector containing the complete information about the track.

The CNN processing aims to exploit the correlations between the LKr, MUV1 and MUV2 signals. Most CNN architectures require an arbitrary number of matrices of identical dimensions (known in the ML context as {\it channels}) as input. Therefore the LKr, MUV1 and MUV2 matrices are transformed to a fixed size of $44\times 44$ by replicating the element values (though the LKr and MUV matrices correspond to different transverse areas). The resulting matrices are illustrated for two events in Fig.~\ref{fig:event_display}. The inputs are finally arranged into a $5\times 44 \times 44$ input tensor.

\section{Particle identification algorithm}
\label{sec:calopid_algo}
\begin{figure}
\resizebox{\textwidth}{!}{%
\begin{tikzpicture}[%
    >=triangle 60,             
    start chain=going right,  
    node distance=20mm and 6mm, 
    every join/.style={->,draw},
    ]
    
\tikzset{
base/.style={draw, on chain, on grid, align=center, minimum height=4ex},
proc/.style={base, rectangle, text width=8em},
test/.style={base, diamond, aspect=2, text width=8em},
term/.style={proc, rounded corners},
coord/.style={coordinate, on chain, on grid, node distance=20mm and 6mm},
}

\node [proc, densely dotted] (p0) {Track};
\node [test, join] (t0) {$E/p$ and $p$ cuts}; 
\node [test] (t1) {MUV3 signal?};
\node [proc] (p1) {Filter};
\node [test,join] (t2) {MIP?};
\node [proc] (p2) {ML model};
\node [term,join, below=of p2] (p3) {$\left(p_\mu,p_\pi,p_e\right)$};
\node[term,below=of t0] (p4) {Stop};

\node [coord, below=of t1] (c1)  {}; 
\node [coord, below=of t2] (c2)  {}; 
\node [coord, below=of p3] (c3)  {}; 

\node [left=1 mm of p3,yshift=1em] {$p_\mu = 1$};

\path (t0.east) to node [near start, yshift=1em] {yes} (t1);
  \draw [o->] (t0.east) -- (t1);
\path (t1.east) to node [near start, yshift=1em] {no} (p1);
  \draw [o->] (t1.east) -- (p1);
\path (t2.east) to node [near start, yshift=1em] {no} (p2); 
  \draw [o->] (t2.east) -- (p2); 
\path (t0.south) to node [near start, xshift=1em] {no} (p4);
  \draw [*->] (t0.south) -- (p4); 
\path (t1.south) to node [near start, xshift=1em] {yes} (c1); 
  \draw [*-] (t1.south) -- (c1) -- (c2);
\path (t2.south) to node [near start, xshift=1em] {yes} (c2); 
  \draw [*->] (t2.south) -- (c2) -- (p3);
\end{tikzpicture}
}
\caption{Flow chart of the developed calorimetric PID algorithms.}
\label{fig:overall_model_flow_chart}
\end{figure}
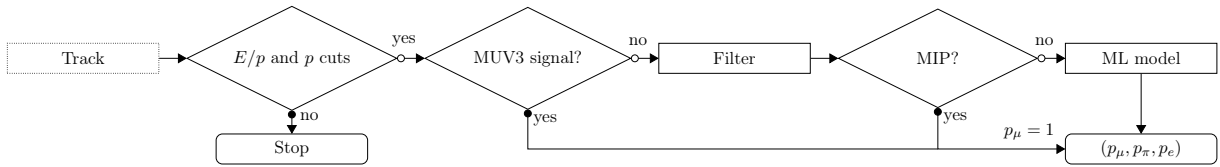

The flow of the developed PID algorithm is illustrated in Fig.~\ref{fig:overall_model_flow_chart}.
The algorithm returns the probabilities $p_\pi$, $p_\mu$, $p_e$ of the track to be classified as $\pi^+$, $\mu^+$, $e^+$, respectively. Tracks with LKr energy deposit to momentum ratio $E/p<0.95$ are considered, which reduces the $e^+$ contamination in the $\pi^+$ and $\mu^+$ samples.
Tracks with matching MUV3 signals, or identified as minimum-ionising particles (MIPs) by a dedicated filter, are assigned $p_\mu=1$. For the remaining tracks, a ML model (either a GBM or a CNN algorithm) is employed to evaluate the probabilities $p_\pi$, $p_\mu$, $p_e$. The models are trained for tracks in the momentum range 15--40~GeV/$c$.

Two MIP filter algorithms, which use the same input information as the ML models, are considered. The filter~A identifies MIPs as tracks with at most four geometrically associated in-time signals in LKr calorimeter cells, at most two hits in each MUV1 view, and at most one hit in each MUV2 view. The filter~B additionally exploits the fact that muons typically produce narrow calorimeter signal patterns. Both ML models have been tested with each of the two filters; the corresponding setups are referred to as GBM/A, GBM/B, CNN/A and CNN/B.

\begin{figure}
\centering
\resizebox{!}{0.5\linewidth}{%
\begin{tikzpicture}[%
    >=triangle 60,
    start chain=going below, 
    node distance=8mm and 6mm,
        every join/.style={norm},
    ]

\tikzset{
  base/.style={draw, on chain, on grid, align=center, minimum height=4ex},
  proc/.style={base, rectangle, text width=16em,rounded corners},
  inpt/.style={base, ellipse,minimum width=16em},
  coord/.style={coordinate, on chain, on grid, node distance=0mm and 40mm},
  add/.style={base, circle},
  norm/.style={->, draw},
}

\node [inpt] (p1) {Input tensor};
\node [proc, join] (p2) {$7 \times 7$ convolution layer};
\node [proc, join]  (p3)    {Batch normalisation};
\node [proc, join] (p4)    {ReLU};
\node [proc, join] (p5)    {ResBlock $\times 2$};

\node [proc, join] (p7)    {ResBlock $\times 2$};
\node [proc, join] (p8)    {ResBlock $\times 2$};
\node [proc, join] (p9)    {Global average pooling};
\node [proc, join] (p10)    {Fully connected layer};
\node [proc, join] (p11)    {Softmax};

\path (p1.south) to node [ right, xshift=1em ] {$5 \times 44 \times 44$} (p2);
\path (p2.south) to node [ right, xshift=1em ] {$64 \times 22 \times 22$} (p3);
\path (p3.south) to node [ right, xshift=1em ] {$64 \times 22 \times 22$} (p4);
\path (p4.south) to node [ right, xshift=1em ] {$64 \times 22 \times 22$} (p5);
\path (p5.south) to node [ right, xshift=1em ] {$64 \times 22 \times 22$} (p7);
\path (p7.south) to node [ right, xshift=1em ] {$128 \times 11 \times 11$} (p8);
\path (p8.south) to node [ right, xshift=1em ] {$256 \times 6 \times 6$} (p9);
\path (p9.south) to node [ right, xshift=1em ] {$256 \times 1 \times 1$} (p10);
\path (p10.south) to node [ right, xshift=1em ] {$3$} (p11);
\node [left=of p2] (c5)  {$S = 2$}; 
\node [left=of p7] (c3)  {$S = 2$}; 
\node [left=of p8] (c4)  {$S = 2$}; 
\end{tikzpicture}
}

\caption{The NA62ResNet architecture. Residual blocks (ResBlock) are stacked in pairs. The stride $S$~\cite{Murphy2022} is a parameter of the filter. The output tensor dimensions at each step are indicated.}
\label{fig:NA62_Resnet_arc}
\end{figure}
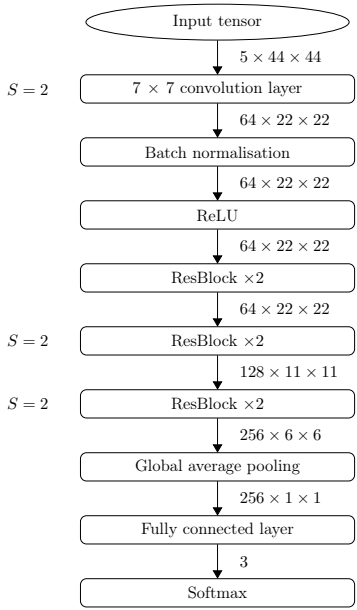

The GBM implementation uses the LightGBM framework, which is efficient for large datasets and high-dimensional feature spaces. A CNN architecture of the ResNet-18 type~\cite{he2016deep} is used.
To handle the sparsely populated input matrices (Fig.~\ref{fig:event_display}), the $3\times3$ maximum pooling layer after the first batch normalisation layer is removed. The network is further simplified by removing the last two residual blocks that include four 512-input convolution layers. The output of the last of the remaining residual blocks is down-sampled via a global average pooling layer, which is followed by a fully connected layer producing the final output subsequently normalised using a softmax function to obtain the probabilities $p_\pi$, $p_\mu$ and $p_e$. The resulting NA62ResNet architecture is shown in Fig.~\ref{fig:NA62_Resnet_arc}.

Data augmentation is used to improve the robustness and generalisation of the CNN model: for each epoch of the training, the five input matrices are mirror-imaged with respect to the horizontal axis for 50\% of the tracks in the training sample chosen at random. The weights and the bias terms of the fully connected layer are initialised with random numbers distributed uniformly in the  $(-1/\sqrt{N},1/\sqrt{N})$ range, where $N=256$ is the number of input features.
The convolution layers are initialised using the He scheme~\cite{He2015}, and the cross entropy loss function is used. The network parameters are optimised following the Adam method~\cite{kingma2014adam}. To detect potential overfitting, the validation loss is monitored during training. The PyTorch framework~\cite{Paszke2019} is used to build and train the model. The architecture is integrated into the NA62 software framework.

The numbers of $\pi^+$, $\mu^+$ and $e^+$ tracks in the training and validation dataset passing the MIP filter~A and used for training (Fig.~\ref{fig:overall_model_flow_chart})
are $6.9 \times 10^6$, $1.0 \times 10^7$ and $7.8 \times 10^{4}$, respectively. 
During the model training, 25\% of the dataset is randomly set aside to form a validation sample, and the rest is allocated to the training sample. The cross entropy loss function~\cite{Murphy2022} is used for training of both GBM and CNN models. 

\section{Performance of the algorithm}
\label{sec:results}

\begin{figure}[p]
\centering
\qquad
\subfloat{{\includegraphics[width=0.5\linewidth]{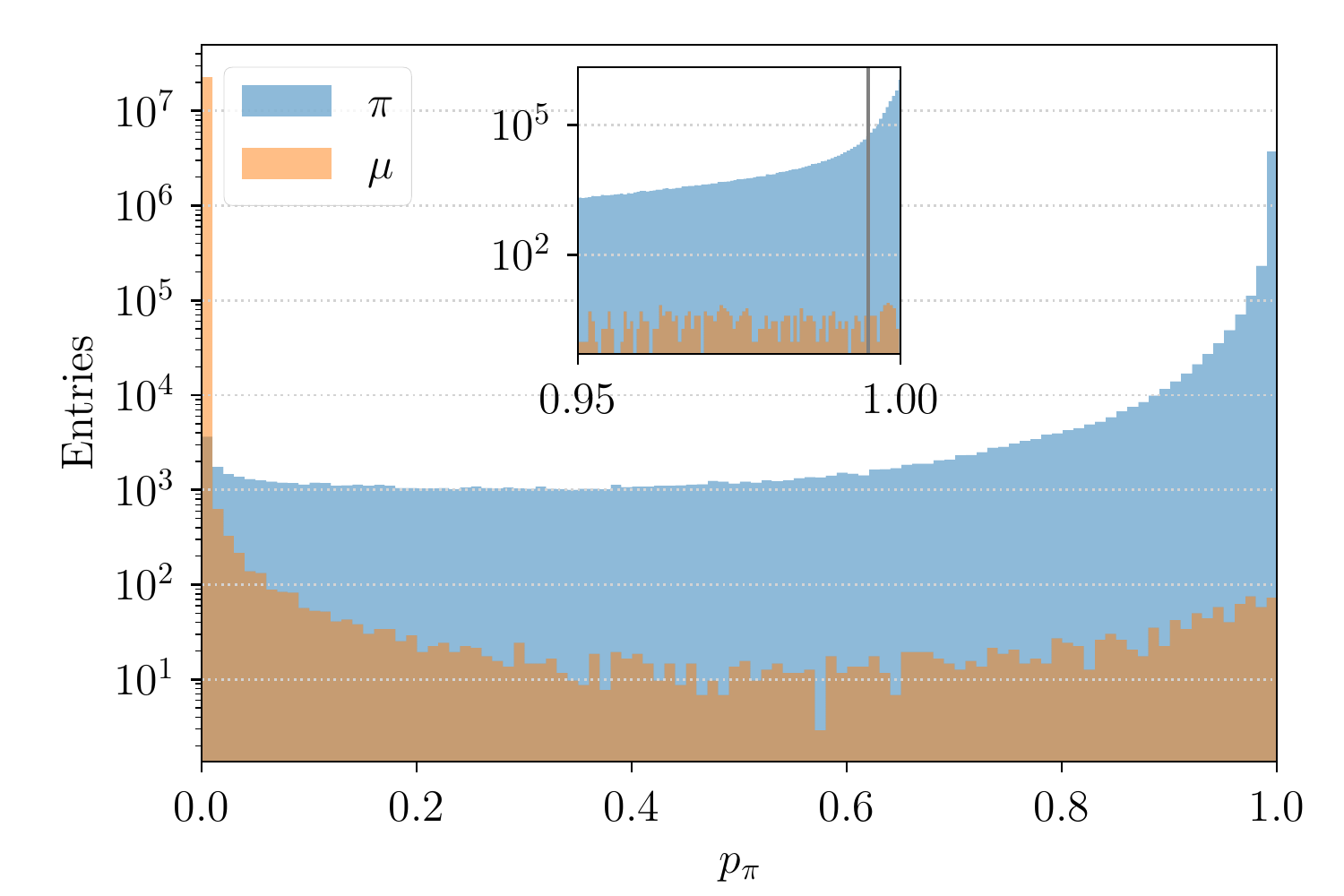}}}%
\subfloat{{\includegraphics[width=0.5\linewidth]{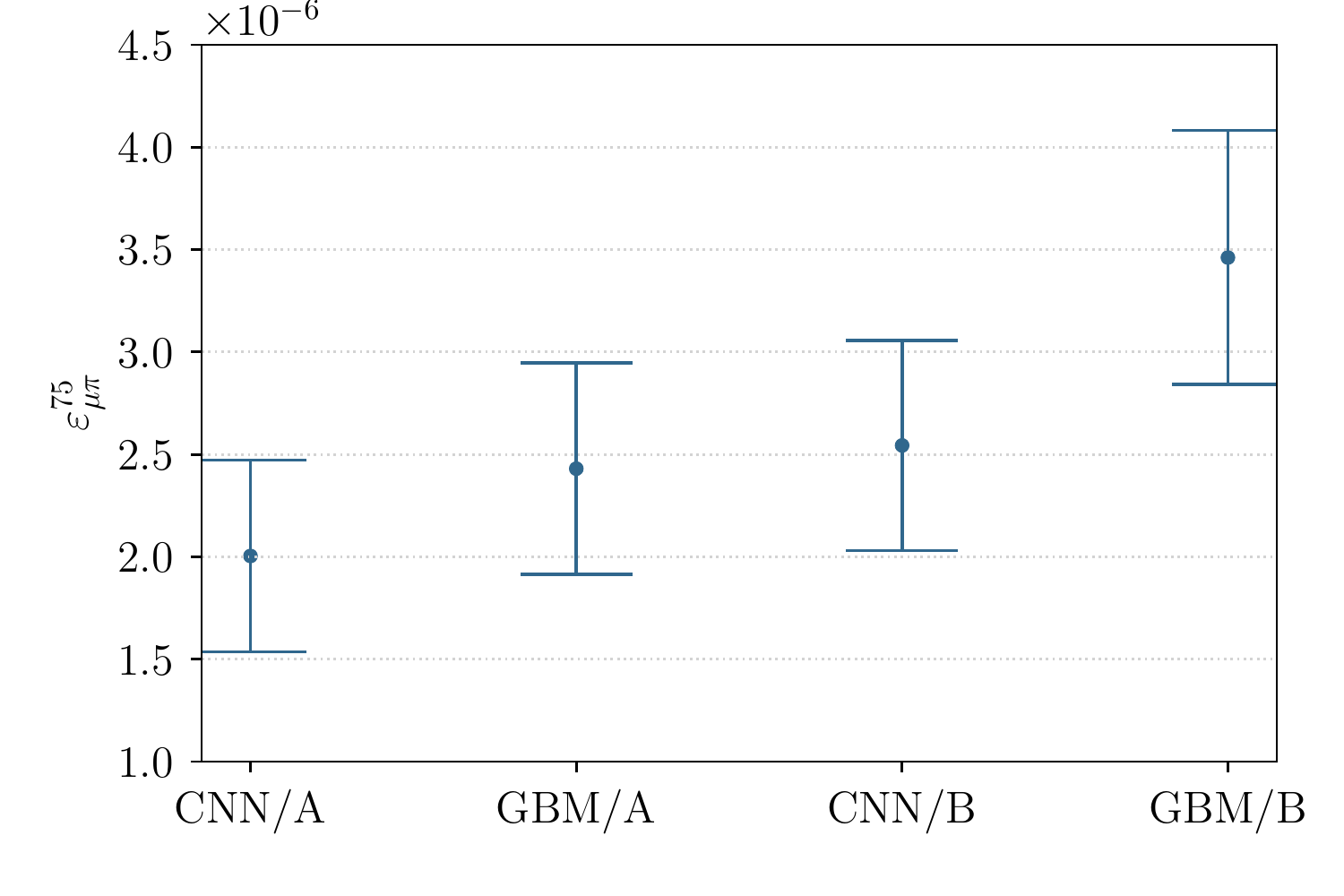}}}%
\caption{Left: $p_\pi$ distributions for $\pi^+$ and $\mu^+$ tracks in the momentum range of 15--40~GeV/$c$ obtained with the CNN/A setup for the test dataset. Inset: the high-$p_\pi$ region; the threshold $p_0$ chosen for $\pi^+$ identification is indicated with a vertical line. Right: $\mu^+$ misidentification probabilities ($\varepsilon_{\mu\pi}^{75}$) with their statistical uncertainties, measured for each setup using a subset of the test dataset.}
\label{fig:muon-misid}
\end{figure}

\begin{figure}[p]
\centering
\qquad
\subfloat{{\includegraphics[width=0.5\linewidth]{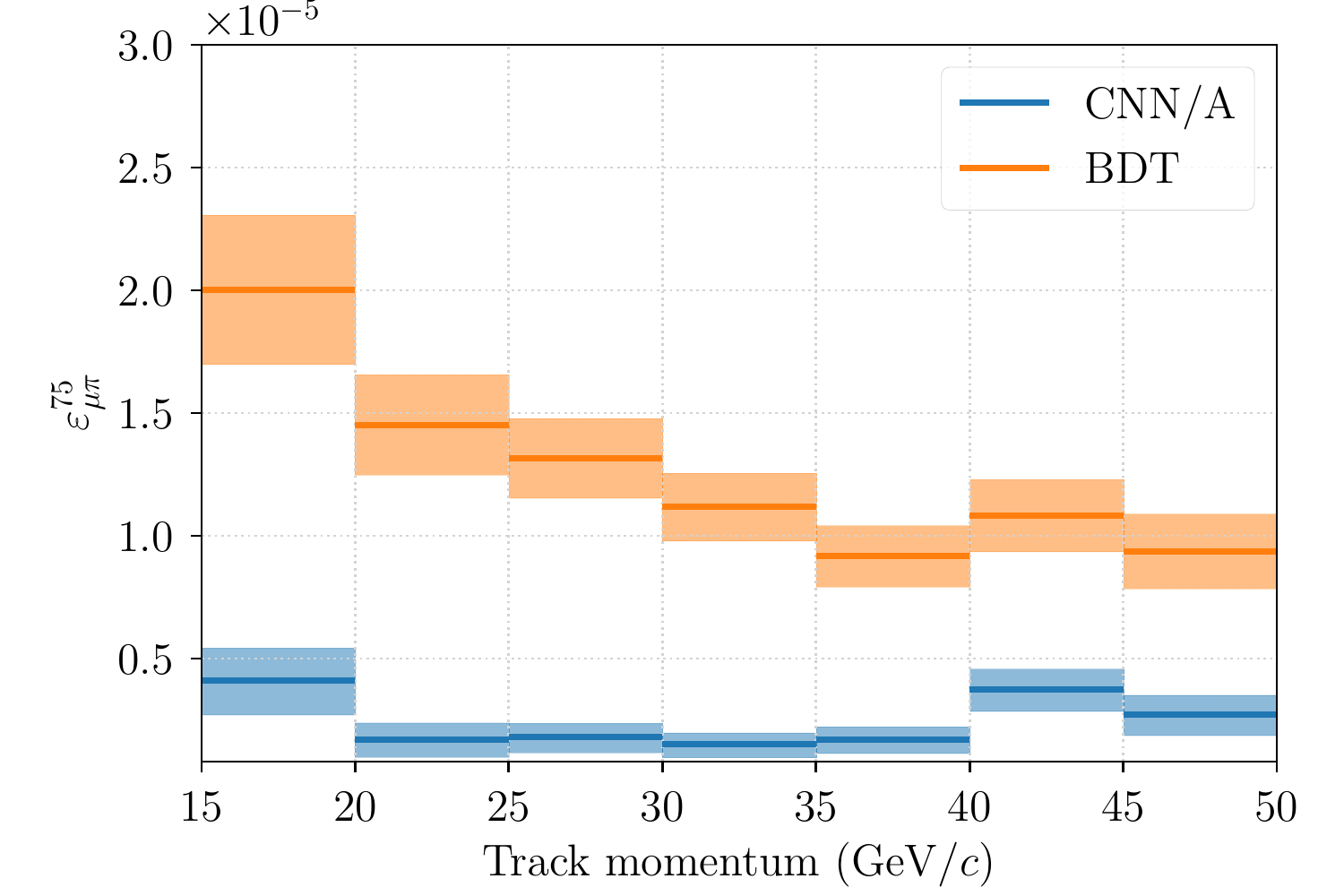}}}%
\subfloat{{\includegraphics[width=0.5\linewidth]{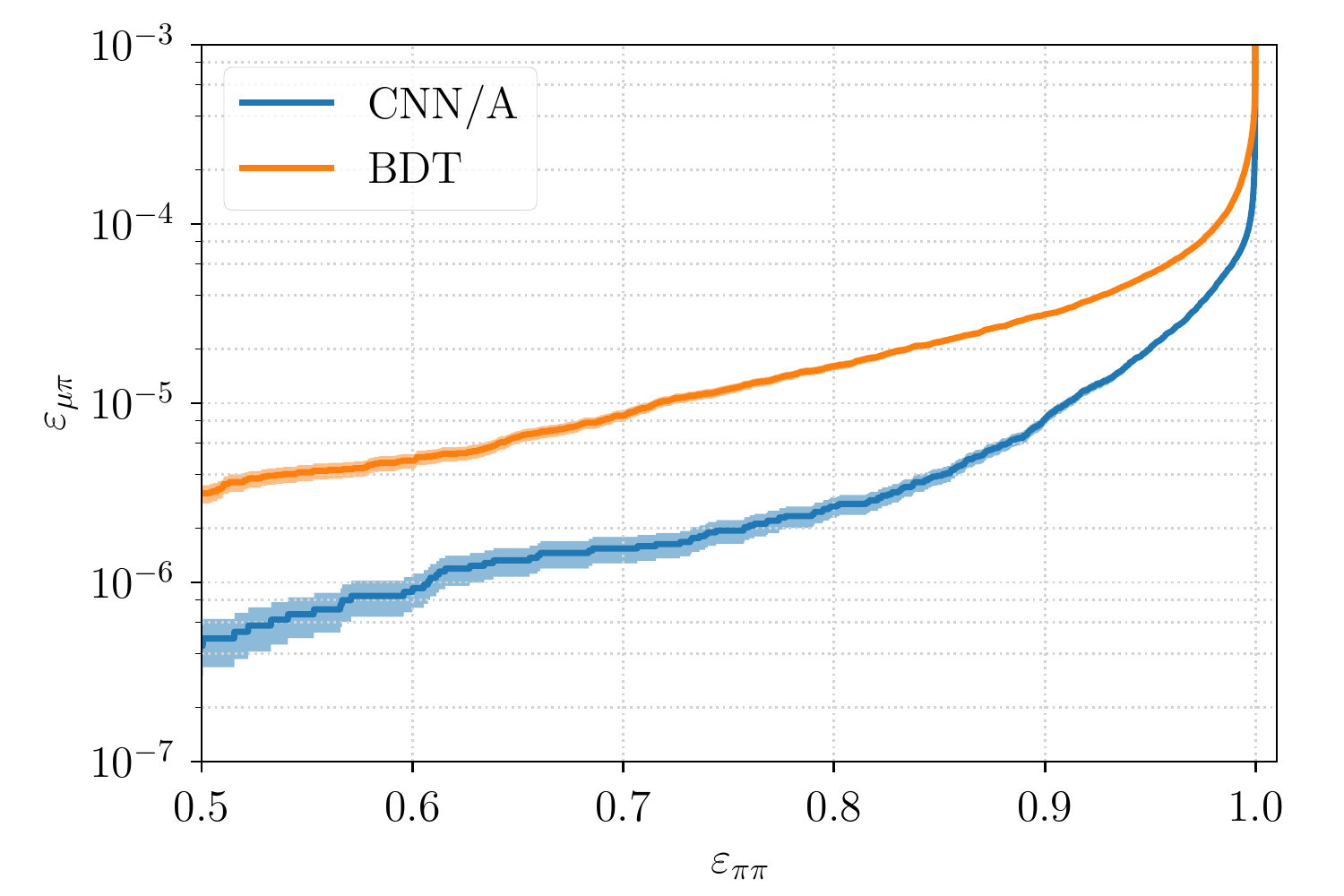}}}%
\caption{Left: muon misidentification probability ($\varepsilon_{\mu\pi}^{75}$) as a function of track momentum, measured with the CNN/A setup for the test dataset. Right: the ROC curve, i.e. muon misidentification probability ($\varepsilon_{\mu\pi}$) vs pion identification efficiency ($\varepsilon_{\pi\pi}$) for a set of $p_0$ values, obtained with the CNN/A setup for the test dataset in the 15--40~GeV/$c$ momentum range. The performance of the BDT algorithm currently used by NA62 is also shown in both panels. The shaded areas represent the statistical uncertainties.}
\label{fig:P_mu_pi}
\end{figure}

The results presented in this section are obtained using the test dataset. Background contamination in the $\pi^+$ and $\mu^+$ samples in the dataset is found using Monte Carlo simulations to be below $10^{-5}$ and $10^{-6}$, respectively. The $p_\pi$ distributions for $\pi^+$ and $\mu^+$ tracks obtained with the CNN/A setup for tracks in the momentum range of 15--40~GeV/$c$ are shown in Fig.~\ref{fig:muon-misid}~(left). A condition $p_\pi>p_0$ is used for $\pi^+$ identification, where $p_0$ is chosen for each setup to obtain a $\pi^+$ identification efficiency of 75\%. The corresponding $\mu^+$ misidentification probabilities, $\varepsilon_{\mu\pi}^{75}$, measured for each setup using a subset of data are displayed in Fig.~\ref{fig:muon-misid}~(right). The CNN/A setup is found to provide the lowest $\varepsilon_{\mu\pi}^{75}$ value, and is used for further analysis. The results reported below are obtained with the CNN/A setup.

The measured $\varepsilon_{\mu\pi}^{75}$ value as a function of track momentum is shown in Fig.~\ref{fig:P_mu_pi}~(left). Strong $\mu^+$ suppression  is achieved also in the momentum range 40--50~GeV/$c$ not used for training. The receiver operating characteristic (ROC) curve in the momentum range 15--40~GeV/$c$ is shown in Fig.~\ref{fig:P_mu_pi}~(right). An improvement in $\varepsilon_{\mu\pi}^{75}$ by a factor of six, from $1.2\times 10^{-5}$ to $2.0\times 10^{-6}$, is obtained with respect to the BDT algorithm currently used by NA62. Alternatively, the $\pi^+$ identification efficiency in the above momentum range is increased from 72\% to 91\% when the muon misidentification probability $\varepsilon_{\mu\pi}$ is kept at a fixed level of $10^{-5}$.
The $e^+$ misidentification probability ($\varepsilon_{e\pi}^{75}$) is 35\% lower in comparison to the BDT algorithm.

Training the CNN/A model separately in each momentum bin leads to an increase of $\varepsilon_{\mu\pi}^{75}$ by a factor of two across the momentum bins in comparison with Fig.~\ref{fig:P_mu_pi}, which is attributed to the smaller size and $\pi^+/\mu^+$ imbalance of the training and validation datasets in the individual bins. Training the model  in the entire 15--50~GeV/$c$ range increases $\varepsilon_{\mu\pi}^{75}$ by about 20\% across the momentum bins in comparison with  Fig.~\ref{fig:P_mu_pi}, which is attributed to $e^+$ contamination in the training and validation dataset above 40~GeV/$c$.

\section{Summary}
A new calorimetric particle identification algorithm based on a convolutional neural network classifier augmented by a filter has been developed for the NA62 experiment at CERN. With respect to the BDT algorithm currently used by NA62, muon misidentification probability as a pion in the momentum range 15--40~GeV/$c$ is reduced by a factor of six from \mbox{$1.2\times 10^{-5}$} to \mbox{$2.0\times 10^{-6}$}, for a fixed pion-identification efficiency of 75\%. Alternatively, pion identification efficiency is improved in the above momentum range from 72\% to 91\% for a fixed muon misidentification probability of $10^{-5}$. 
The algorithm is applicable to a wide range of NA62 physics analyses, with best performance achieved after careful parameter tuning.

\section*{Acknowledgements}
The authors are grateful to T.~Timbers and the University of British Columbia Master of Data Science Capstone team for their assistance. The authors would also like to thank W.~Deng, B.~Lie and S.~Sethi for their contributions to an early phase of this work. 
It is a pleasure to express our appreciation to the staff of the CERN laboratory and the technical
staff of the participating laboratories and universities for their efforts in the operation of the
experiment and data processing.

The cost of the experiment and its auxiliary systems was supported by the funding agencies of 
the Collaboration Institutes. We are particularly indebted to: 
F.R.S.-FNRS (Fonds de la Recherche Scientifique - FNRS), under Grants No. 4.4512.10, 1.B.258.20, Belgium;
CECI (Consortium des Equipements de Calcul Intensif), funded by the Fonds de la Recherche Scientifique de Belgique (F.R.S.-FNRS) under Grant No. 2.5020.11 and by the Walloon Region, Belgium;
NSERC (Natural Sciences and Engineering Research Council), funding SAPPJ-2018-0017,  Canada;
MEYS (Ministry of Education, Youth and Sports) funding LM 2018104, Czech Republic;
BMBF (Bundesministerium f\"{u}r Bildung und Forschung) contracts 05H12UM5, 05H15UMCNA and 05H18UMCNA, Germany;
INFN  (Istituto Nazionale di Fisica Nucleare),  Italy;
MIUR (Ministero dell'Istruzione, dell'Universit\`a e della Ricerca),  Italy;
CONACyT  (Consejo Nacional de Ciencia y Tecnolog\'{i}a),  Mexico;
IFA (Institute of Atomic Physics) Romanian 
CERN-RO No. 1/16.03.2016 
and Nucleus Programme PN 19 06 01 04,  Romania;
MESRS  (Ministry of Education, Science, Research and Sport), Slovakia; 
CERN (European Organization for Nuclear Research), Switzerland; 
STFC (Science and Technology Facilities Council), United Kingdom;
NSF (National Science Foundation) Award Numbers 1506088 and 1806430,  U.S.A.;
ERC (European Research Council)  ``UniversaLepto'' advanced grant 268062, ``KaonLepton'' starting grant 336581, Europe.

Individuals have received support from:
Charles University Research Center (UNCE/SCI/013), Czech Republic;
Ministero dell'Istruzione, dell'Universit\`a e della Ricerca (MIUR  ``Futuro in ricerca 2012''  grant RBFR12JF2Z, Project GAP), Italy;
the Royal Society  (grants UF100308, UF0758946), United Kingdom;
STFC (Rutherford fellowships ST/J00412X/1, ST/M005798/1), United Kingdom;
ERC (grants 268062,  336581 and  starting grant 802836 ``AxScale'');
EU Horizon 2020 (Marie Sk\l{}odowska-Curie grants 701386, 754496, 842407, 893101, 101023808).

\printbibliography

\end{document}